\documentclass[preprint,sort&compress,5p]{elsarticle}
\usepackage{graphicx}
\usepackage{amsmath,amssymb}
\usepackage[nodots]{numcompress}

\journal{Physics Letters B}

\begin{document}

\begin{frontmatter}
\title{The importance of being warm (during inflation)}

\author[a]{Sam Bartrum}
\ead{sam.bartrum@ed.ac.uk}

\address[a]{SUPA, School of Physics and Astronomy, University of Edinburgh, 
Edinburgh, EH9 3JZ, United Kingdom}

\author[b]{Mar Bastero-Gil}
\ead{mbg@ugr.es}

\address[b]{Departamento de F\'{\i}sica Te\'orica y del Cosmos, 
Universidad de Granada, Granada-18071, Spain}

\author[a]{Arjun Berera}
\ead{ab@ph.ed.ac.uk}

\author[b]{Rafael Cerezo}
\ead{cerezo@ugr.es}

\author[c]{Rudnei O. Ramos}
\ead{rudnei@uerj.br}
\address[c]{Departamento de F\'isica Te\'orica, Universidade do Estado do 
Rio de Janeiro, 
20550-013 Rio de Janeiro, RJ, Brazil}

\author[d]{Jo\~ao G. Rosa\corref{e}}
\ead{joao.rosa@ua.pt} 
\address[d]{Departamento de F\'{\i}sica da Universidade de Aveiro and I3N, Campus de Santiago, 3810-183 Aveiro, Portugal} 

\cortext[e]{Corresponding author}

\begin{abstract}

The amplitude of primordial curvature perturbations is enhanced when a radiation bath at a temperature $T>H$ is sustained during inflation by dissipative particle production, which is particularly significant when a non-trivial statistical ensemble of inflaton fluctuations is also maintained. Since gravitational modes are oblivious to dissipative dynamics, this generically lowers the tensor-to-scalar ratio and yields a modified consistency relation for warm inflation, as well as changing the tilt of the scalar spectrum. We show that this alters the landscape of observationally allowed inflationary models, with for example the quartic chaotic potential being in very good agreement with the Planck results for nearly-thermal inflaton fluctuations, whilst essentially ruled out for an underlying vacuum state. We also discuss other simple models that are in agreement with the Planck data within a renormalizable model of warm inflation.
\end{abstract}

\begin{keyword}
warm inflation; chaotic inflation; Cosmic Microwave Background; arXiv:1307.5868 [hep-ph]
\end{keyword}

%\pacs{98.80.Cq, 11.30.Pb, 12.60.Jv} 

%\preprint{UG-FT 305/13, CAFPE 175/13, Edinburgh 2013/19}

\end{frontmatter}

%%%%%%%%%%%%%%%%%%%%%%%%%%%%%%%%%%%%%%%%%%%%%%%%%%%%%%%%%%%%%%%%%%%%%%%%%%%%%%%%%%%%%%%%%%%%%%%%%%%%%%%%%%%%%%%%%%%%%%%%%%%%%%%%%%%%%%%%%%%%%%%%%%%%%%%%%%%%%%%%%%

\section{Introduction}

More than three decades since its original proposal \cite{inflation}, inflation has passed one of its most stringent tests with the recent measurements of the temperature anisotropies of the Cosmic Microwave Background (CMB) made by the Planck satellite \cite{Ade:2013uln}. In particular, models of inflation based on the dynamics of a slowly rolling scalar field generate a primordial spectrum of density perturbations that is essentially adiabatic, gaussian and nearly but not exactly scale-invariant, in agreement with observations. 

While Planck has been able to place strong constraints on several inflationary models, the question remains as to which is the fundamental mechanism driving inflation. Ca-nonical models assume that accelerated expansion quickly erases all traces of any pre-inflationary matter or radiation distributions, so that slow-roll inflation occurs in an almost perfect vacuum state. However, the inflaton field is necessarily coupled to other degrees of freedom in order to dissipate its vacuum energy and reheat the universe, so one may envisage scenarios where dissipative effects become important during and not only after the slow-roll phase. These are generically known as warm inflation scenarios \cite{Berera:1995wh, Berera:2008ar}, and raise the interesting possibility that the inflationary universe is not in a perfect vacuum, even though vacuum energy is the dominant component for accelerated expansion to take place. 

In this Letter, we show that even when dissipative effects are still small compared to Hubble damping, the amplitude of scalar curvature fluctuations can be significantly enhanced, whereas tensor perturbations are  generically unaffected due to their weak coupling to matter fields. This generically reduces the tensor-to-scalar ratio with respect to conventional models and also modifies the tilt of the scalar power spectrum, thereby changing observational predictions considerably. These effects are particularly significant when the temperature of the radiation bath $T\gtrsim H$ and also when non-trivial inflaton occupation numbers are sustained during inflation. As an example, we show that the simplest model of chaotic inflation, $V(\phi)=\lambda\phi^4$, falls well within Planck's observational window for a nearly-thermalized state, in a supersymmetric realization of warm inflation with renormalizable interactions, whereas it seems to be ruled out in a cold scenario. 

%%%%%%%%%%%%%%%%%%%%%%%%%%%%%%%%%%%%%%%%%%%%%%%%%%%%%%%%%%%%%%%%%%%%%%%%%%%%%%%%%%%%%%%%%%%%%%%%%%%%%%%%%%%%%%%%%%%%%%%%%%%%%%%%%%%%%%%%%%%%%%%%%%%%%%%%%%%%%%%%%%

\section{Warm inflation dynamics}

The simplest models of warm inflation consider an adiabatic regime, where dissipative processes occur faster than Hubble expansion and the inflaton's slow motion, such that the dissipative part of the inflaton's self-energy from its coupling to other fields leads to an effective friction coefficient in its equation of motion \cite{Moss:1985wn, Yokoyama:1987an, Berera:1995wh, Berera:2008ar}:
\begin{equation} \label{inflaton_eq}
\ddot\phi+3H\dot\phi+\Upsilon\dot\phi+V'(\phi)=0~,
\end{equation}
which reduces to $3H(1+Q)\dot\phi\simeq -V'(\phi)$ in the overdamped or slow-roll regime, where $Q=\Upsilon/3H$. This implies the generalized conditions $\epsilon,|\eta|\ll1+Q$ on the slow-roll parameters $\epsilon=(M_P^2/2)(V'/V)^2$ and $\eta=M_P^2 V''/V$, which in particular allow for $\epsilon, |\eta|\gtrsim 1$ in the strong dissipation regime, $Q\gtrsim 1$, therefore alleviating the need for very flat potentials \cite{Berera:1995wh}. Moreover, the inflaton dissipates its energy through particle production and, if the resulting particles are relativistic and relax to an equilibrium configuration sufficiently fast, a nearly-thermal radiation bath is sourced all through inflation: 
\begin{equation} \label{radiation_eq}
\dot\rho_R+4H\rho_R=\Upsilon\dot\phi^2~,
\end{equation}
with $\rho_R=(\pi^2/30)g_*T^4$ for $g_*$ relativistic degrees of freedom. In the slow-roll regime, the radiation quickly reaches a quasi-stationary configuration where $\dot\rho_R$ may be neglected and we obtain $\rho_R/\rho_\phi\simeq (\epsilon/2)Q/(1+Q)^2\ll1$. Radiation is then a sub-dominant component, allowing for accelerated expansion, although one may have $T>H$ even for weak dissipation, $Q\ll1$. At the end of inflation, however, for $\epsilon\sim 1+Q$ we have $\rho_R/\rho_\phi\sim (1/2)Q/(1+Q)$, and radiation may become a relevant component if $Q\gtrsim 1$ at this stage. In fact, in models where dissipation becomes stronger as inflation proceeds, radiation will typically come to dominate once the slow-roll regime has ended, yielding a smooth `graceful exit' into a radiation-dominated universe.

A well-studied dissipation mechanism considers a supersymmetric model with a renormalizable superpotential of the form \cite{Berera:2002sp, Moss:2006gt}:
\begin{equation} \label{superpotential}
W=f(\Phi)+{g\over2}\Phi X^2+ {h\over 2}XY^2~,
\end{equation}
where the inflaton corresponds to the scalar component of the chiral multiplet $\Phi$, $\phi=\sqrt{2}\langle\Phi\rangle$, with a scalar potential $V(\phi)=|f'(\phi)^2|$ that spontaneously breaks supersymmetry (SUSY) during inflation. Dissipation is sourced by the coupling of the inflaton to the bosonic and fermionic $X$ fields and their subsequent decay into $Y$ scalars and fermions, which form the thermal bath. In the low temperature regime, $m_X\simeq (g/\sqrt{2})\phi\gtrsim T$, which typically corresponds to large field values and where inflaton thermal mass corrections are Boltzmann-suppressed \cite{highT}, dissipation is mainly mediated by the decay of virtual scalar modes $\chi\rightarrow y y$, yielding \cite{BasteroGil:2012cm, Moss:2006gt}:
\begin{equation} \label{dissipation_coefficient}
\Upsilon=C_\phi {T^3\over \phi^2}~,\qquad C_\phi\simeq  {1\over4}\alpha_h N_X~,
\end{equation}
for $\alpha_h=h^2N_Y/4\pi\lesssim1$ and $N_{X,Y}$ chiral multiplets. Supersymmetry suppresses radiative and thermal corrections to the scalar potential, yielding at 1-loop order:
\begin{equation} \label{1-loop}
 \Delta V^{(1)}/V\simeq (\alpha_g/8\pi)\log(m_X^2/\mu^2)~,
\end{equation}
where $\alpha_g=g^2N_X/4\pi\lesssim 1$ and $\mu$ is the renormalization scale.

%inflaton particle production
The heavy fields also decay into inflaton particle states through $\chi\rightarrow yy\phi$, but this is a sub-leading process, with $\Gamma(\chi\rightarrow yy\phi)= (g/4\pi)^2\Gamma(\chi\rightarrow yy)$, where $\Gamma(\chi\rightarrow yy)=\alpha_h m_X/16$ \cite{BasteroGil:2012cm}, with inflaton particles from this process alone typically yielding a negligible component of the radiation bath. However, dissipative particle production destabilizes the local thermal equilibrium of the plasma, triggering decays, inverse decays and thermal scatterings that redistribute the dissipated energy between all the interacting fields and keep the system close to equilibrium if occurring faster than the Hubble rate. In particular, decays and inverse decays can be efficient thermalization processes \cite{Anisimov:2008dz} so that, in some parametric regimes, we expect inflaton particles to be sustained in a quasi-thermal state at the ambient temperature $T$ by decays and inverse decays of the multiple heavy species in the plasma.

Although the details of the thermalization process require solving the system of coupled Boltzmann equations for all the particle species involved, which may require full numerical simulations and is outside the main scope of this Letter, the underlying physical picture can be understood in simple terms. Starting from an equilibrium configuration where decay and inverse decay processes are occurring at equal rates, dissipation of the inflaton's energy will mainly produce an excess of light particles in the $Y$ sector, which will enhance the rate of inverse decays and consequently increase the $X$ sector occupation numbers above their equilibrium value. This in turn enhances the direct decay rate, producing $Y$ particles and also an excess of inflaton modes. This goes on until the balance between decay and inverse decay rates is restored and the system reaches a new equilibrium configuration. One then expects the energy injected into the system to be distributed amongst all species in the plasma that are produced and annihilated faster than Hubble expansion. The common temperature of these species would decrease due to expansion but this is compensated by the additional energy, keeping the temperature roughly constant in the slow-roll regime. Species that are not created/destroyed sufficiently fast will decouple from the plasma and their effective temperature will be exponentially redshifted away during inflation, quickly reaching a quasi-vacuum state. A measure of the efficiency of the thermalization processes can then be obtained by comparing the relevant decay rates with the Hubble parameter, as we examine in more detail below in the context of chaotic inflation.

%%%%%%%%%%%%%%%%%%%%%%%%%%%%%%%%%%%%%%%%%%%%%%%%%%%%%%%%%%%%%%%%%%%%%%%%%%%%%%%%%%%%%%%%%%%%%%%%%%%%%%%%%%%%%%%%%%%%%%%%%%%%%%%%%%%%%%%%%%%%%%%%%%%%%%%%%%%%%%%%%% 
 
\section{Primordial power spectrum}

Fluctuation-dissipation dynamics modifies the evolution of inflaton perturbations, which are sourced by a gaussian white noise term, $\xi_k$ \cite{Berera:1995wh, Berera:1999ws, Hall:2003zp}:
\begin{equation} \label{fluctuation_eq}
\delta\ddot\phi_k+3H(1+Q)\delta\dot\phi_k+{k^2\over a^2}\delta\phi_k\simeq \sqrt{2\Upsilon T}a^{-3/2}\xi_k~,
\end{equation}
in the slow-roll regime. We recall that the intensity of the noise is related to the dissipation coefficient through the fluctuation-dissipation theorem and, as shown in the first article in \cite{BasteroGil:2012cm}, dissipation from scalar modes dominates over the fermionic one in the low temperature regime that we are considering here. Spatial correlation properties may, however, be different for fermionic modes \cite{Yamaguchi:1996dp}. Also, as mentioned above, dissipative processes may maintain a non-trivial distribution of inflaton particles, $n(k)$, which for sufficiently fast interactions should approach the Bose-Einstein distribution at the ambient temperature, $n_{BE}(k)=(e^{k/aT}-1)^{-1}$. In this case the associated creation and annihilation operators have correlation functions $\langle\hat{a}_{-\mathbf{k}}\hat{a}_\mathbf{k'}^\dagger\rangle=[n+1](2\pi)^3\delta^3(\mathbf{k}+\mathbf{k'})$ and $\langle\hat{a}_\mathbf{k'}^\dagger\hat{a}_{-\mathbf{k}}\rangle=n(2\pi)^3\delta^3(\mathbf{k}+\mathbf{k'})$. For a generic inflaton phase-space distribution at the time when observable CMB scales leave the horizon during inflation, $t_*$, one then obtains the dimensionless power spectrum \cite{Berera:1995wh, Moss:1985wn, Berera:1999ws, Hall:2003zp, Ramos:2013nsa}:
\begin{equation} \label{power_spectrum}
\Delta_\mathcal{R}^2=\left({H_*\over\dot\phi_*}\right)^2\left({H_*\over 2\pi}\right)^2\left[1+2n_*+\left(T_*\over H_*\right){2\sqrt{3}\pi Q_*\over\sqrt{3+4\pi Q_*}}\right]~,
\end{equation}
which yields the standard cold inflation result in the limit $n_*, Q_*, T_*\rightarrow 0$. This expression neglects, however, the coupling between inflaton and radiation fluctuations associated with the temperature dependence of the dissipation coefficient in Eq.~(\ref{dissipation_coefficient}), an effect that may significantly enhance the perturbation growth for strong dissipation, $Q\gtrsim 1$ \cite{Graham:2009bf}. Since this coupling is negligible if the relevant scales become super-horizon when dissipation is weak, we can obtain an accurate description of the spectrum by taking the limit $Q_*\ll 1$ in the expression above, which yields:
\begin{equation} \label{scalar_power_spectrum_full}
\Delta_\mathcal{R}^2\simeq\left({H_*\over\dot\phi_*}\right)^2\left({H_*\over 2\pi}\right)^2\left[1+2n_*+2\pi Q_*{T_*\over H_*}\right]~.
\end{equation}
This assumption is justified in models where $Q$ grows during inflation, such that dissipation has a negligible effect 50-60 e-folds before the end of inflation but becomes stronger towards the end, thus helping to prolong the period of accelerated expansion. We show below that this is indeed the case for our working example of chaotic inflation and discuss the potential effects of strong dissipation at horizon-crossing at the end of this Letter.

Note that both the second and third terms within the brackets  in Eq.~(\ref{scalar_power_spectrum_full}) are positive-definite, the former corresponding to non-trivial inflaton occupation numbers and the latter to the leading effect of fluctuation-dissipation dynamics. Hence, the amplitude of the scalar power spectrum always exceeds the vacuum result in warm inflation scenarios. On the other hand, gravity waves are weakly coupled to the thermal bath and the spectrum of tensor modes retains its vacuum form, $\Delta_t^2=(2/\pi^2)(H_*^2/ M_P^2)$. This therefore suppresses the tensor-to-scalar ratio, yielding a modified consistency relation for warm inflation: 
\begin{equation} \label{consistency}
r\simeq{8|n_t|\over 1+2n_*+2\pi Q_*T_*/H_*}~,
\end{equation}
where $n_t=-2\epsilon_*$ is the tensor index. The primordial tensor spectrum can thus be used to distinguish warm from cold inflation scenarios, the former consequently evading the Lyth bound \cite{Lyth:1996im, Cai:2010wt} (see also \cite{Hotchkiss:2011gz} for other scenarios where the Lyth bound does not apply). Whereas previous studies have focused more on the strong dissipation regime, this result explicitly shows that warm inflation can have a significant observational impact for weak dissipation, where temperatures well above the Hubble rate can be sustained. Most importantly, non-trivial inflaton occupation numbers may also generically lower the tensor-to-scalar ratio, which as we illustrate below may have a very significant effect on inflationary predictions.

In the limit where inflaton particle production is inefficient and $n_*$ gives a negligible contribution to the power spectrum, the scalar spectral index is nevertheless modified by the third term in Eq.~(\ref{scalar_power_spectrum_full}), yielding:
\begin{equation} \label{scalar_index_vacuum}
n_s-1\simeq 2\eta_{*}-6\epsilon_{*}+\frac{2\kappa_*}{1+\kappa_*}\left(7\epsilon_{*}-4\eta_{*}+5\sigma_{*}\right)~,
\end{equation}
where $\sigma=M_P^2 V'/(\phi V)<1+Q$ and we have used the slow-roll equations, $3H(1+Q)\dot\phi\simeq -V'(\phi)$ and $\rho_R\simeq (3/4)Q\dot\phi^2$, to determine the variation of $\kappa\equiv2\pi QT/H$ as different scales become super-horizon during inflation.

Modifications are, however, more prominent in the opposite limit of nearly-thermal inflaton fluctuations, with $n_*\simeq n_{BE*}$. For $T_*\gtrsim H_*$ and $Q_*\ll1$ we then obtain:
\begin{equation} \label{scalar_index_thermal}
n_s-1\simeq 2\sigma_{*}-2\epsilon_{*}~,
\end{equation}
which is, in particular, independent of the curvature of the potential, which only determines its running:
\begin{equation} \label{running}
n_s'\simeq 2\sigma_{*}(\sigma_{*}+2\epsilon_{*}-\eta_{*})-4\epsilon_{*}(2\epsilon_{*}-\eta_{*})~.
\end{equation}
In this case, a red-tilted spectrum, $n_s<1$, corresponds to either potentials with a negative slope, such as hill-top models, or large field models where $\epsilon_{*}> 2(M_P/\phi_*)^2$. 

%%%%%%%%%%%%%%%%%%%%%%%%%%%%%%%%%%%%%%%%%%%%%%%%%%%%%%%%%%%%%%%%%%%%%%%%%%%%%%%%%%%%%%%%%%%%%%%%%%%%%%%%%%%%%%%%%%%%%%%%%%%%%%%%%%%%%%%%%%%%%%%%%%%%%%%%%%%%%%%%%% 
 
\section{Chaotic warm inflation}

To illustrate the effects of both dissipation and occupation numbers on observational predictions, let us consider the quartic model, $V(\phi)=\lambda\phi^4$, which corresponds to a superpotential $f(\Phi)=\sqrt{\lambda}\Phi^3/3$ and is the canonical model of chaotic inflation \cite{Linde:1983gd}. In this case, we have $\epsilon=2\sigma= 2\eta/3=8(M_P/\phi)^2$, which yields $n_s-1\simeq -8(M_P/\phi_*)^2$ for a thermalized inflaton distribution from Eq.~(\ref{scalar_index_thermal}). This gives a red-tilted spectrum with $n_s\simeq 0.96$ for $\phi_*\simeq 14 M_P$, which is super-planckian but smaller than the corresponding field value in the vacuum case, $\phi_*\simeq 25 M_P$. This also gives $r\simeq 8(1-n_s)(H_*/T_*)$, within the upper bound obtained by Planck, $r<0.11$ (95\% CL), for $T_*> 2.9H_*$, as well as a small negative running $n_s'=-(n_s-1)^2\simeq -0.0016$ and a tensor index $n_t=2(n_s-1)\simeq-0.079$.

%e-folds
The number of e-folds of inflation can be computed by integrating the slow-roll equations, which may be done analytically for the quartic model \cite{Bartrum:2012tg}. 
In particular, one can use the form of the dissipation coefficient in Eq.~(\ref{dissipation_coefficient}) to express the coupled inflaton and radiation equations in the slow-roll regime as a single equation for the dissipative ratio $Q$:
\begin{equation} \label{Q_eq}
{dQ\over dN_e}= C_*{Q^{6/5}(1+Q)^{6/5}\over 1+7Q}~,
\end{equation}
where $C_*\simeq 5\epsilon_*Q_*^{-1/5}$ for $Q_*\ll1$. This shows explicitly that $Q$ grows during inflation, justifying our assumption that the system may evolve from the weak to the strong dissipation regime. Inflation ends in this case when $|\eta|=1+Q$, which yields $Q_e\simeq (2/3(1-n_s))^{5/2}Q_*^{1/2}$ for a thermal spectrum and hence $Q_e\gtrsim 1$ for $Q_*\gtrsim 10^{-6}$. As discussed earlier, the relative abundance of radiation will then also grow towards the end of inflation, with $\rho_R/V(\phi)\propto Q^{7/5}$ in this case, until it smoothly takes over after slow-roll has ended. Integrating Eq.~(\ref{Q_eq}) from horizon-crossing to the end of the slow-roll regime, we obtain:
\begin{equation} \label{e-folds}
N_e\simeq\epsilon_{*}^{-1}\left(1+bQ_*^{1/5}\right)~,
\end{equation}
where $b\simeq 2.81$. This yields the required $50-60$ e-folds of inflation with $n_s\simeq0.96-0.97$ for $Q_*\simeq 0.001-0.01$. For comparison, in the standard cold inflation regime, one finds $n_s=1-3/N_e$, giving $n_s=0.94-0.95$ for $N_e=50-60$. This clearly shows that even for weak dissipation at horizon-crossing one may obtain substantially different observational results.

For both limits of nearly-thermal and negligible inflaton occupation numbers, one can use the observed amplitude of curvature perturbations,  $\Delta_\mathcal{R}^2\simeq 2.2\times 10^{-9}$ \cite{Ade:2013uln} and the form of the dissipation coefficient in Eq.~(\ref{dissipation_coefficient}) to relate the different quantities at horizon-crossing, with e.g.~$Q_*\simeq 2\times10^{-8} g_* (H_*/T_*)^{3}$ in the nearly-thermalized regime. This allows one to express both $n_s$ and $r$ in terms of the dissipative ratio or temperature at horizon-crossing for a given number of e-folds of inflation and relativistic degrees of freedom, which is illustrated in Figure 1.

As one can see, observational predictions for the quartic model depend on the distribution of inflaton fluctuations, $n_*$. For $n_*, \kappa_*\ll1$, the spectrum has the same form as in cold inflation, but from Eq.~(\ref{e-folds}) one obtains $N_e=50-60$ for smaller field values than in cold inflation, yielding a larger tensor fraction and a more red-tilted spectrum. When $\kappa_*\gtrsim 1$, however, the spectrum becomes more blue-tilted and $r$ is suppressed, although for weak dissipation it remains too large.

%%%%%%%%%%%%%%%%%%%%
%%%%%%%%%%%%%%%%%%%%
\begin{figure} [htbp]
\centering\includegraphics[scale=0.33]{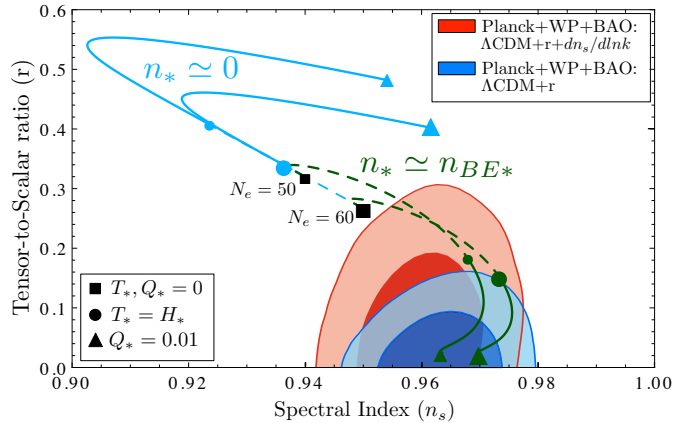}
\caption{Trajectories in the $(n_s,r)$ plane for $V(\phi)=\lambda\phi^4$ as a function of the dissipative ratio, $Q_*<0.01$, 50-60 e-folds before the end of inflation, compared with the Planck results  \cite{Ade:2013uln}, for $g_*=228.75$ relativistic degrees of freedom. The dark green (light blue) curves correspond to nearly-thermal (negligible) inflaton occupation numbers $n_*$, with dashed branches for $T_*\lesssim H_*$. Note that corresponding curves converge in the cold inflation limit, $T_*,Q_*\rightarrow 0$.}
\end{figure}
%%%%%%%%%%%%%%%%%%%%
%%%%%%%%%%%%%%%%%%%%

On the other hand, for nearly-thermal inflaton occupation numbers tensor modes are more strongly suppressed and one obtains a remarkable agreement with the Planck results for $T_*\gtrsim H_*$. Note that for $T_*\lesssim H_*$ the concept of thermal equilibrium is ill-defined, since the average particle modes have super-horizon wavelengths, so in Figure 1 we represent this regime with dashed curves to nevertheless illustrate the transition from a cold to a warm spectrum. Also, we take the MSSM value $g_*=228.75$ only as a reference, with fewer light species further lowering the tensor-to-scalar ratio, since $T_*/H_*$ is larger.

%chaotic inflation
This agreement is particularly significant, since the quartic potential is the simplest renormalizable model of chaotic inflation, involving no other scales other than the inflaton field value. As originally argued by Linde \cite{Linde:1983gd}, in large-field models inflation is naturally triggered from a chaotic field distribution following the pre-planckian era, in domains where $V(\phi)\sim M_P^4$ quickly dominates over gradient and kinetic energy densities. On the other hand, when inflation only occurs for a $V(\phi)\ll M_P^4$ plateau, the post-planckian universe must be unnaturally smooth, requiring a fine-tuning of initial conditions that the inflationary paradigm is supposed to solve \cite{Ijjas:2013vea}.  

While other modifications such as a non-minimal coupling to gravity may also bring the quartic model into agreement with observations \cite{Kallosh:2013pby}, the renormalizable nature of the interactions leading to dissipation is an attractive feature of warm inflation, with only a few controllable parameters. Note, in particular, that interactions with other bosonic and/or fermionic fields are always required since the vacuum energy of the inflaton field must be transferred into light degrees of freedom at the end of inflation to `reheat' the universe. In this sense, warm inflation scenarios do not introduce any non-standard modifications to the basic inflationary models but simply correspond to parametric regimes where the universe is kept warm throughout inflation, $T\gtrsim H$. For the dissipation coefficient in Eq.~(\ref{dissipation_coefficient}), one obtains in particular $T_*/H_*\sim (C_\phi/g_*) N_e^{-2}\gtrsim 1$, which may be achieved for $N_X\gg N_Y\gtrsim 1$ and $g, h\ll 1$, while keeping radiative corrections under control,  $\alpha_{g,h}\lesssim 1$. We may express the number of heavy species as: 
\begin{eqnarray} \label{number_heavy_fields}
N_X \simeq {8\times 10^5\over \alpha_h} \left({0.04\over 1-n_s}\right)^4\left({r\over 0.01}\right)^2\left({Q_*\over 10^{-3}}\right)~,
\end{eqnarray}
where we have assumed a thermal distribution of inflaton perturbations. This large multiplicity of $X$ species is typical of the form of the dissipation coefficient in Eq.~(\ref{dissipation_coefficient}) \cite{BasteroGil:2009ec, Cerezo:2012ub}, but is expected to be significantly reduced in other regimes, such as for on-shell $X$ modes \cite{BasteroGil:2012cm}. Large multiplicities may be obtained in D-brane constructions \cite{BasteroGil:2011mr}, where the $X$ fields correspond to strings stretched between brane and antibrane stacks and their number thus grows with the square of the brane multiplicity. Due to brane-antibrane annihilation at the end of inflation, these modes will not, however, play a role in the post-inflationary universe. Field multiplicities are also enhanced by the Kaluza-Klein tower in extra-dimensional scenarios \cite{Matsuda:2012kc}.

%warm baryogenesis
An interesting possibility arises when we consider B- and CP-violating interactions for the $X$ fields in Eq.~(\ref{superpotential}), with complex couplings and distinct decay channels. In this case, the out-of-equilibrium nature of dissipation can generate a cosmological baryon asymmetry during inflation \cite{BasteroGil:2011cx}. The resulting baryon-to-entropy ratio depends on the inflaton field, so that inflaton fluctuations yield both adiabatic and baryon isocurvature (BI) perturbations with a nearly-scale invariant spectrum. For the quartic model with $n_*\simeq n_{BE*}$, BI and adiabatic modes are anti-correlated with relative amplitude $B_B\simeq 3(n_s-1)\simeq-0.12$ and a blue-tilted spectrum $n_{iso}\simeq (3-n_s)/2\simeq 1.02$ \cite{Bartrum:2013oka}. This then yields for the relative matter isocurvature spectrum $\beta_{iso}\simeq (\Omega_b/\Omega_c)^2B_B^2\simeq4.8\times10^{-4}$, well within the bound $\beta_{iso}<0.0087$ obtained by Planck for anti-correlated isocurvature modes with $n_s\simeq n_{iso}$, which is in fact the case that best improves the fit to the data \cite{Ade:2013uln}.

The interactions required to produce a baryon asymmetry through dissipation are analogous to those considered in conventional thermal GUT baryogenesis or leptogenesis models, with the scalar $X$ fields corresponding to e.g.~heavy GUT bosons or right-handed neutrinos \cite{BasteroGil:2011cx}. However, since only virtual $X$ modes are involved in the dissipative processes, baryogenesis may occur below the GUT scale, as opposed to thermal GUT baryogenesis models, avoiding the production of dangerous relics such as monopoles. In particular, we obtain for the temperature at the end of inflation in the quartic model:
\begin{equation} \label{temperature_end}
T_e\simeq 10^{14}\left({1-n_s\over 0.04}\right)^{5\over2}\left({0.01\over r}\right)^{1\over2}\left({10^{-3}\over Q_*}\right)^{3\over10}\ \mathrm{GeV}~.
\end{equation}
We note that the effective reheating temperature is roughly an order of magnitude lower since radiation typically takes a few e-folds to take over after the end of slow-roll \cite{Bartrum:2012tg}. While gravitinos may still be ubiquitously produced at these temperatures, the inflaton may not decay completely right after inflation if $Q_e\lesssim 10$  \cite{Sanchez:2010vj}, as is the case of the quartic model for $Q_*<0.01$. The inflaton may then come to dominate over the radiation bath at a later stage and the entropy produced  by its eventual decay may dilute the excess of gravitinos, thus avoiding the potentially associated cosmological problems \cite{Sanchez:2010vj}.

%thermalization
Our results motivate a closer look at thermalization processes and, in particular, we can estimate the total production rate of inflaton particles from the 3-body decay rate of the $N_X$ heavy species in the plasma given above. At horizon-crossing, the inflaton is a relativistic degree of freedom, since $m_{\phi_*}=\sqrt{3}\eta_*H_*\ll T_*$, and we obtain:
\begin{equation} \label{inflaton_production}
{\Gamma_{\phi_*}\over H_*}\simeq 9\left(\alpha_g\alpha_h\right)^{3\over2}\left({1-n_s\over 0.04}\right)^{3\over2}\left({0.01\over r}\right)^{3\over2}\left({0.005\over Q_*}\right)^{1\over2},
\end{equation}
where we assumed $n_*=n_{BE*}$. Moreover, finite temperature Bose factors may considerably enhance this for small couplings \cite{Drewes:2013iaa}, with e.g.~the two-body decay width increasing up to a factor  $T/m_Y\sim \sqrt{12}/h$ \cite{BasteroGil:2012cm}. Also, $\Gamma_\phi/H$ increases during inflation, so that deviations from thermal equilibrium should become less significant. We then expect inflaton particles to be produced sufficiently fast and remain close to thermal equilibrium with the ambient plasma if the effective couplings $\alpha_{g,h}$ are not too small. Both the inflaton and other light fields could actually be in a pre-inflationary thermal state with $T\gtrsim H$, with dissipation and the above mentioned processes maintaining a slowly-varying temperature. Without dissipation, however, thermal effects would be quickly redshifted away, yielding quite different observational features \cite{Bhattacharya:2005wn}. 

%%%%%%%%%%%%%%%%%%%%%%%%%%%%%%%%%%%%%%%%%%%%%%%%%%%%%%%%%%%%%%%%%%%%%%%%%%%%%%%%%
%%%%%%%%%%%%%%%%%%%%%%%%%%%%%%%%%%%%%%%%%%%%%%%%%%%%%%%%%%%%%%%%%%%%%%%%%%%%%%%%%

\section{Conclusion}

In this Letter, we have shown that the presence of even small dissipative effects at the time when observable scales leave the horizon during inflation may have a significant effect on the spectrum of primordial fluctuations in the warm regime, for $T\gtrsim H$. This generically lowers the tensor-to-scalar ratio and yields a modified consistency relation for warm inflation that may be used to distinguish it in a model-independent way from the standard supercooled scenarios if a tensor component is found and accurately measured. The main modifications to the scalar spectrum arise from the presence of dissipative noise that sources inflaton fluctuations and from the changes in the phase space distribution of inflaton modes as a consequence of inflaton particle production in the plasma. We have shown, in particular, that the latter effect may bring the simplest chaotic inflation scenario with a quartic potential into agreement with the Planck results for a nearly-thermal distribution. Inflation may thus be triggered from chaotic initial conditions at the Planck scale in an observationally consistent way, through simple renormalizable interactions with matter fields that must be present in any inflationary model, as opposed to e.g.~a non-minimal coupling to the gravitational sector. The cosmic baryon asymmetry may also be produced during warm inflation, inducing baryon isocurvature perturbations that are within the current Planck bounds for a quartic potential and which may be probed in the near future.

%other potentials
Although for the quartic model a nearly-thermal spectrum is observationally preferred, as e.g. for $V(\phi)\propto \phi^6$, this is not necessarily true in general. For example, models such as SUSY hybrid inflation driven by small radiative corrections \cite{Dvali:1994ms}, which follows from Eq.~(\ref{superpotential}), are consistent with the Planck data when $n_*$ is negligible and only the fluctuation-dissipation term modifies the spectrum. In particular, dissipation increases the number of e-folds in this case, whereas in the cold regime only $N_e<50$ is observationally allowed \cite{Ade:2013uln}. Other low-scale models such as hill-top scenarios are consistent for both the thermal regime \cite{BuenoSanchez:2008nc} and when the fluctuation-dissipation term in Eq.~(\ref{scalar_power_spectrum_full}) is dominant, the same holding for exponential \cite{Yokoyama:1987an} or inverse power-law potentials, although an alternative reheating mechanism is needed for $Q_*\ll1$ since dissipation never becomes sufficiently strong in this case. A detailed analysis of these and other potentials is, however, outside the scope of this Letter and will be presented elsewhere.

%strong dissipation
We have focused on the regime where dissipation is still sub-leading at horizon-crossing, which is simpler to realize since it requires smaller values of $T_*/H_*$ and lower field multiplicities. Non-gaussian effects should also be suppressed in this case, with $f_{NL}\sim \mathcal{O}(1)$ by extrapolating the results in \cite{Moss:2007cv} to weak dissipation,  within the bounds obtained by Planck \cite{Ade:2013uln}. The Planck collaboration has searched for signals of non-gaussianity in warm inflation models in the strong dissipation regime \cite{Ade:2013ydc}, but a dedicated analysis of the bispectrum for $Q_*\ll1$ is required and further motivated by our results. 

For strong dissipation the coupling between inflaton and radiation perturbations enhances the growth of fluctuations, making the spectrum more blue-tilted for increasing $Q$. To check the validity of our results for the quartic potential, we have evolved the fluctuations numerically extending the analysis in \cite{BasteroGil:2011xd} to weak dissipation, the details of which will be presented in a separate publication. We find that the primordial spectrum is modified if $Q_*\gtrsim 0.01$ when the largest scales become super-horizon, since in this case the smallest observable scales leave the horizon 8-10 e-folds later when $Q\gtrsim1$. Shear viscous effects have been shown to suppress the enhanced growth \cite{BasteroGil:2011xd}, but since these imply departures from thermal equilibrium a more detailed analysis is required. The results presented in Figure 1 for $Q_*<0.01$ are nevertheless in good agreement with the numerical spectrum. Note that even weak dissipative effects can lower the tensor-to-scalar ratio below the expected reach of Planck, $r\gtrsim 0.05$, or other CMB polarization experiments such as the Atacama B-mode Search or the South Pole Telescope, with $r\gtrsim 0.03$.

The significant changes to the primordial spectrum from dissipative and thermal effects may also extend beyond the SUSY realization of warm inflation considered in this Letter. In non-SUSY models, for example, while couplings between the inflaton and other fields must be smaller to prevent large radiative corrections, there may exist parametric regimes where thermalization occurs sufficiently fast to yield near-equilibrium occupation numbers. One could also envisage alternative models, e.g.~the decay of multiple scalar fields that become underdamped at different stages during inflation and produce inflaton particles during the first few oscillations about the minima of their potential. 

Warm inflation may thus provide a first principles dynamical mechanism to sustain non-trivial occupation numbers for the inflaton and other light particles, based on renormalizable interactions,  and it would be interesting to investigate whether phenomenological excited states \cite{Landau:2011aa, Ashoorioon:2013eia} could find concrete realizations in this context. The most important effect of dissipation and/or a non-trivial inflaton particle distribution is the lowering of the tensor-to-scalar ratio in the modified consistency relation in Eq. (\ref{consistency}), so we expect next year's Planck release and future CMB B-mode polarization searches to shed new light on the nature of inflaton fluctuations.

%%%%%%%%%%%%%%%%%%%%%%%%%%%%%%%%%%%%%%%%%%%%%%%%%%%%%%%%%%%%%%%%%%%%%%%%%%%%%%%%%%%%%%%%%%%%%%%%%%%%%%%%%%%%%%%%%%%%%%%%%%%%%%%%%%%%%%%%%%%%%%%%%%%%%%%%%%%%%%%%%%%%%%%%%%%%%%%%%%%%%%%%%%%%%%%%%%%%%%%%%%%%%%%%%%%%%%%%%%%%%%%%%%%%%%%%%%%%%%%%%%%%%%%%%%%%%%%%%%%%%%%%%%%%%%%%%%

\section*{Acknowledgements}

SB and AB are supported by STFC. MBG and RC are partially supported by MICINN (FIS2010-17395) and ``Junta de Andaluc\'ia'' (FQM101).  ROR is
partially supported by CNPq and FAPERJ (Brazil). JGR is supported by FCT (SFRH/BPD/85969/2012) and partially by the grant PTDC/FIS/116625/2010 and the Marie Curie action NRHEP-295189-FP7-PEOPLE-2011-IRSES.

%%%%%%%%%%%%%%%%%%%%%%%%%%%%%%%%%%%%%%%%%%%%%%%%%%%%%%%%%%%%%%%%%%%%%%%%%%%%%%%%%%%%%%%%%%%%%%%%%%%%%%%%%%%%%%%%%%%%%%%%%%%%%%%%%%%%%%%%%%%%%%%%%%%%%%%%%%%%%%%%%%%%%%%%%%%%%%%%%%%%%%%%%%%%%%%%%%%%%%%%%%%%%%%%%%%%%%%%%%%%%%%%%%%%%%%%%%%%%%%%%%%%%%%%%%%%%%%%%%%%%%%%%%%%%%%%%%
\bibliographystyle{model1-num-names}

\end{document}